\documentclass[preprint]{elsarticle}
\usepackage{setspace}
\usepackage{graphicx}
\usepackage{natbib}
\usepackage{amsmath, amssymb}
\usepackage[colorlinks,urlcolor=blue,citecolor=blue,linkcolor=blue]{hyperref}

\newcommand{\aj}{AJ}
\newcommand{\araa}{ARA\&A}
\newcommand{\apj}{ApJ}
\newcommand{\apjl}{ApJ}

\newcommand{\aap}{A\&A}

\newcommand{\mnras}{MNRAS}

\newcommand{\nat}{Nature}

\newcommand{\icarus}{Icarus}

\begin{document}

\title{Consequences of an eccentric orbit for Fomalhaut b}
\author{Daniel Tamayo}
\ead{dtamayo@astro.cornell.edu}
\address{Department of Astronomy, Cornell University, Ithaca, NY 14853}

\begin{abstract}
Fomalhaut b is currently the least massive, directly imaged exoplanet candidate.  New observation epochs reveal this object to be moving on a highly eccentric orbit \citep{Kalas13}, which sets important new constraints.  I consider scenarios where Fomalhaut b is the only object interacting with the debris disk, and ones involving an additional unseen planet.  I also investigate the possibility that Fomalhaut b is merely a transient dust cloud in light of the revised eccentric orbit.  I argue that the scenario best able to match the observational constraints is a super-Earth Fomalhaut b surrounded by a vast cloud of dust that is generated by a population of irregular satellites, with an undetected $\sim$ Saturn-mass planet orbiting interior to the disk and driving the secular dynamics.  Testable predictions are summarized that could differentiate between this scenario and other possibilities.
\end{abstract}

\maketitle

\section{Introduction}
Fomalhaut b is currently the least massive, directly imaged exoplanet candidate \citep{Kalas08}.  \cite{Kalas05} had predicted the existence of such a planet due to their discovery that the system's circumstellar debris disk was eccentric ($e \approx 0.1$).  This was significant since one expects collisional dissipation to circularize the orbits of bodies in the belt of debris.  However, Laplace-Lagrange secular theory shows that planets on elliptical orbits can force similarly eccentric orbits in test particles.  \cite{Quillen06} performed an in-depth study determining that not only could an unseen planet induce the belt's observed eccentricity, it could also explain the disk's observed sharp inner edge as the boundary of the planet's chaotic zone where mean motion resonances overlap and particle orbits are quickly depleted.  It was therefore exciting when, using optical images from the Hubble Space Telescope (HST), \cite{Kalas08} discovered Fomalhaut b at approximately the predicted projected separation from the star in an orbit that seemed capable of explaining the disk's eccentricity \citep{Chiang09}.

However, Fomalhaut b's planetary status has been controversial.  While \cite{Currie12} and \cite{Galicher13} performed independent analyses of the optical HST data that confirmed the detection of Fomalhaut b, searches in the infrared \citep{Kalas08, Marengo09, Janson12} did not recover the putative planet.  This implies that rather than directly imaging the planet, optical images are detecting starlight scattered by a vast dust cloud, which has led some \citep[e.g.,][]{Janson12} to discard the planetary interpretation completely.  In this case, however, one must posit an additional unseen perturbation to force the debris disk's observed eccentricity.

\cite{Kalas08} instead interpreted the dust as lying in an optically thick protosatellite disk about a planet.  In principle, this could simultaneously explain both the observed flux and the debris-disk dynamics; however, it is unclear whether such a disk could persist for the system's age of $440 \pm 40$ Myr \citep{Mamajek12}.  \cite{Kennedy11} instead suggested that a more diffuse and extended circumplanetary cloud of debris would be a natural consequence of the planet hosting a population of irregular satellites---small, captured moons at the outskirts of a planet's sphere of gravitational influence.  Each of the Solar System's four outer planets hosts many such satellites, suggesting they could be a general feature of giant planets \citep[see reviews by][]{Jewitt07, Nicholson08}.  Furthermore, an analogous vast dust ring sourced by the irregular satellite Phoebe persists around Saturn \citep{Verbiscer09}.  While this Saturnian disk is faint, the Solar System's irregular satellite size-distribution suggests that when the Sun was as young as the central star Fomalhaut A, the dust clouds around the giant planets would have been much brighter \citep{Bottke10}, perhaps yielding comparable fluxes to those observed from Fomalhaut b \citep{Kennedy11}.  Thus, a circumplanetary dust cloud sourced by irregular satellites plausibly explains the observations, and a massive planet on a low-eccentricity orbit ($e \approx 0.1$) can account for the otherwise puzzling elliptical circumstellar debris disk.  

However, \cite{Kalas13}'s (hereafter K13) recent analysis of two additional epochs of HST observations complicate the story further.  They find that Fomalhaut b does not move on a low-eccentricity path interior to the circumstellar debris disk as previously thought; instead, the orbit has an extreme eccentricity of $0.8 \pm 0.1$, and crosses the debris disk (in projection).  This renders the previous low-eccentricity Laplace-Lagrange secular analyses \citep{Quillen06, Chiang09} inapplicable to Fomalhaut b.  One must now ask not only why the circumstellar debris disk is eccentric, but also how it has managed to survive the gravitational perturbations of an object on such an extreme orbit.  One would expect such a crossing orbit to both smear the boundaries of the belt and to puff up its vertical extent, through both secular effects and close encounters.  Instead, one observes sharp inner \citep{Kalas05} and outer \citep{Boley12} edges of a disk that is vertically confined to an opening angle of $1.5^{\circ}$ \citep{Kalas05}.  The seeming incongruity between the debris disk's dynamically cold state and Fomalhaut b's highly elliptical orbit may therefore set important constraints on the putative planet's mass and lifetime on its present orbit.  

In order to understand the limits imposed by Fomalhaut b's revised orbit, K13 performed numerical simulations for several possible scenarios.  In this paper, I perform deeper studies of what I consider to be the most important configurations, probing larger sets of initial conditions and pushing toward longer times.  In particular, while K13 focused mostly on orbital timescales, I probe the longer secular timescales on which the orbits of particles in the debris disk can change dramatically.  I find this provides valuable new constraints.  

I organize my investigation around the need to account for the circumstellar debris disk's observed eccentricity.  There are three broad classes of explanation for its elliptical geometry:  interactions with other stars, hydrodynamical instabilities within the disk, and planetary perturbations.

First, the central star Fomalhaut A is part of a wide binary.  As the two stars undergo a close approach when the companion Fomalhaut B crosses pericenter, one would expect an initially circular debris disk to develop tightly wound, eccentric spiral arms that qualitatively resemble the observed disk \citep{Larwood01}.  But as K13 point out, successive pericenter passages would scramble this structure, and the system has undergone many such events over its lifetime.  The companion is also too widely separated to induce secular dynamical variations in the disk, as these would have periods of $\sim 100$ Gyr (K13).  Recently, a second companion star has been discovered \citep{Mamajek13}, though it too is far enough away that its gravitational effect is negligible.  Looking beyond the system, \cite{Deltorn01} searched for past close encounters with a sample of $\approx 20000$ stars whose space motions could be ascertained.  They found an F7V star imparted the strongest perturbation $474^{+20}_{-19}$ Myr ago, with a closest-approach distance of $1.15^{+0.41}_{-0.34}$ pc.  However, this is probably too far to excite the observed eccentricity in the belt, and may have predated the belt itself.  

In the second class, \cite{Lyra13} have recently proposed a hydrodynamical mechanism that obviates the need for an external perturber.  They find that debris disks with small quantities of gas can spontaneously develop narrow eccentric rings through clumping instabilities.  This is a promising possibility, but requires further observational and theoretical investigation.  Currently, only upper limits exist on the gas content in the Fomalhaut belt \citep{Liseau99}, and current simulations do not produce rings that are wide or eccentric enough to match the observed disk.  While the disk could be made up of a large number of narrow, unresolved rings, only a fraction of annuli become elliptical in the simulations of \cite{Lyra13}.  Resolutions to these problems are beyond the scope of this paper so I do not pursue this possibility further.  

Given the above limitations, I investigate the third scenario, namely, that gravitational perturbations from planets are responsible for the belt's elliptical geometry.  I treat two limits.  In Sec.\ \ref{notOnlyPlanet}, I assume that Fomalhaut b is the only massive object interacting with the disk, and perform numerical simulations to ascertain parameters and dynamical histories consistent with the belt's observed eccentricity.  I then consider the alternate possibility of an additional unseen planet dominantly forcing the disk's elliptic geometry.  As \cite{Quillen06} and \cite{Chiang09} showed, such a planet orbiting interior to the disk can plausibly explain the observed disk geometry, and current infrared surveys \citep{Kalas08, Marengo09, Janson12} have only probed down to masses $> 1 M_J$.  Following K13, I dub this hypothetical planet Fomalhaut c.  Assuming this configuration, in Sec.\ \ref{fomc} I investigate how long different-mass Fomalhaut bs could move on their on their present orbit without disrupting the disk, and in Sec.\ \ref{dust} I describe some difficulties for a transient-dust-cloud interpretation of Fomalhaut b in light of the revised eccentric orbit.  I conclude in Sec.\ \ref{LHB} by comparing the relative likelihoods of these three scenarios, and by summarizing testable predictions that could differentiate among them.

\section{If only Fomalhaut b interacts with the disk, what is Fomalhaut b's maximum lifetime on its present orbit?} \label{notOnlyPlanet}

It is important to point out that if Fomalhaut b is solely responsible for the debris disk's eccentricity, it must be $\sim$ Neptune-mass or larger.  This is because Fomalhaut b can only drive the dynamics in the debris disk if its mass is comparable to or larger than the mass in the belt.  Estimates of the disk mass are uncertain, but range from $\sim 3$ Earth-masses ($M_\oplus$) \citep{Chiang09} to $\sim 30 M_\oplus$, though the upper limit could be as large as $\sim 1$ Jupiter mass if there are objects with radii $\gg 1$km embedded in the disk \citep{Wyatt02}.  Following \cite{Quillen06} and \cite{Chiang09}, I assume the disk's self-gravity is negligible, but as observations better constrain the problem, this approximation may need to be reevaluated.  

I focus on the putative planet's effect on the parent bodies in the belt that inject mass at the top of the collisional cascade, thereby generating the dusty disk.  We cannot observe such parent bodies directly, but the visible debris they generate through collisions will approximately inherit their parent bodies' orbits \citep{Chiang09}.  Radiation pressure will subsequently modify the orbits of small dust grains \citep{Burns79}; however, it is reasonable to assume that if Fomalhaut b were to disrupt the cohesion between parent-body orbits, the dust belt would be similarly dispersed.  This greatly simplifies the analysis, since it allows one to only consider only gravitational perturbations, ignoring radiation pressure and other non-gravitational forces.  Furthermore, it means that one can safely ignore collisions, since, by definition, parent bodies undergo a single collision over the system's lifetime.  This was the approach taken by \cite{Quillen06}.  

\cite{Chiang09} tested the above assumption.  In addition to considering large parent bodies, they included the effects of radiation pressure on the dust grains that are generated.  They found that the scattered light profile derived from dust generated by a sharply confined population of parent bodies is smoothed radially.  Nevertheless, \cite{Chiang09} found that the inner edge of the scattered-light profile matches that of the parent-body population to within 5 $\%$ (cf. Figs. 3 and 5 in their work).  This claim was verified observationally with ALMA observations that are instead sensitive to millimeter grains \citep{Boley12}.  These grain sizes are negligibly affected by radiation pressure and are thus excellent tracers of parent bodies.  As expected, \cite{Boley12} find a sharp inner edge at a location consistent with the optical observations.  I therefore conclude that we are justified in only considering parent-body orbits.

\begin{figure}
\includegraphics[width=10cm]{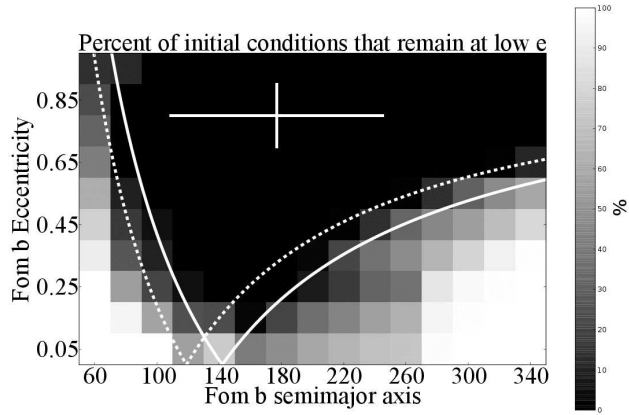}
\caption{\label{percents}  Each grid point, representing a combination of Fomalhaut b's semimajor axis and eccentricity, contains 1000 equally-spaced initial conditions for test particles representing parent bodies in the debris disk.  The gray-scale represents the percentage of those initial conditions that remained at low eccentricity over one secular cycle and are possibly consistent with the observed disk.  The region above the solid white line roughly represents Fomalhaut b orbits that cross the debris disk in projection (the debris disk is assumed circular and at 142 AU).  To within the resolution of our grid, Fomalhaut b orbits that cross the debris disk push nearly all parent bodies onto high-eccentricity orbits.  The region below the dashed white line is inconsistent with the deprojected distance of Fomalhaut b from the central star of 119 AU \citep{Kalas08}.  The white cross gives the marginalized 1-$\sigma$ error bars for Fomalhaut b's orbit from K13. }
\end{figure}

\begin{figure}
\includegraphics[width=10cm]{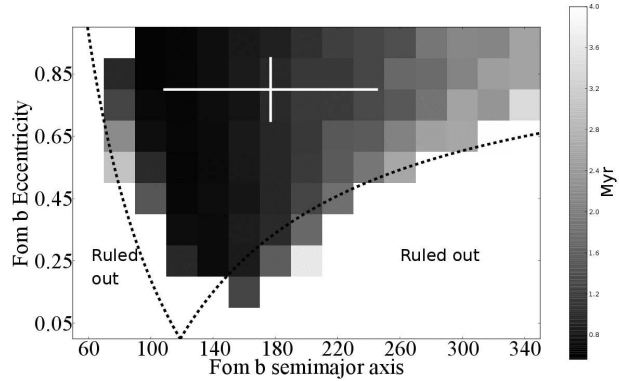}
\caption{\label{times}  Grid and simulations depicted are the same as in Fig.\ \ref{percents}.  The gray-scale now represents the median time required for parent-body orbits in the given grid point to reach $e = 0.3$, assuming Fomalhaut b has Saturn's mass (timescales for other planet masses can be straightforwardly obtained, see text).  Grid points where more than $10\%$ of initial conditions remained at low eccentricity are plotted white.  The region below the two dashed black lines corresponds to the area of phase space that is ruled out by the fact that Fomalhaut b would never reach its current observed distance from the central star of $\approx 119 AU$ \citep{Kalas08}.  The white cross gives the marginalized 1-$\sigma$ error bars for the orbit determination of Fomalhaut b from K13. }
\end{figure}

As for the effects of Fomalhaut b, one might expect that a massive object would disrupt the belt through scattering events during close encounters.  This is true for planetary masses $\gtrsim 1 M_{J}$; for lighter planets, however, the fraction of the belt's circumference that undergoes strong perturbations while Fomalhaut b crosses the disk is small \citep{Tamayo13c}.  K13 also found this in their numerical simulations spanning 2-4$\times 10^5$ yrs (see Fig.\ 29 in their work).  But pushing to longer times, I find that the secular evolution generates a more drastic, global effect.  On a timescale of $M_\star/M_p$ parent-body orbital periods, test-particle orbits undergo large-amplitude eccentricity oscillations approaching unity.  I therefore focus on this slower, secular evolution.  I now describe the simulations and initial conditions in detail.  Readers only interested in the results can skip ahead to Figs.\ \ref{percents} and \ref{times}.

A challenge in dynamically modeling the system with Fomalhaut b's orbit only roughly constrained is that the phase space of initial conditions to sample is large.  But by focusing on the secular evolution, we additionally benefit from a reduced dimensionality of this space.  A classic result by Gauss is that, to first order in the masses, the secular problem is equivalent to smearing out the orbiters' masses along their respective paths in proportion to the time spent at each longitude.  The secular problem can thus be formulated as the interaction between the Fomalhaut b ``mass ring" and the ``rings" corresponding to each of the parent bodies.  In this process, one integrates out the problem's dependence on the orbiters' longitudes.  Furthermore, each average is done over a closed orbit in a time-independent, conservative potential; thus, each object's orbital energy, and therefore each of the semimajor axes, is conserved.  

Considering only Fomalhaut b and an individual parent body, I have reduced the phase space of initial conditions by two by ignoring their respective longitudes.  Furthermore, since the semimajor axes are constant, I can consider their constant ratio as a single parameter.  A final simplification obviates the need to sample a range of planetary masses.  Since I only treat Fomalhaut b as massive, and work to linear order in this mass, I factor out the planet's mass from the perturbing Hamiltonian $\mathcal{H}_p$ acting on each test particle,
\begin{equation}
\mathcal{H}_p = M_p \mathcal{H}'_p,
\end{equation}
where $\mathcal{H}_p$ is the perturbing Hamiltonian, and $\mathcal{H}'_p$ is the ``Hamiltonian" with $M_p$, the mass of Fomalhaut b, factored out.  The change of the orbital elements, expressed in canonical variables, is of the standard Hamiltonian form
\begin{equation}
\frac{d\rho}{dt} = \pm M_p \frac{\partial \mathcal{H}'_p}{\partial \sigma},
\end{equation}
where $\rho$ is a canonical coordinate or momentum (this choice determines the appropriate sign), and $\sigma$ is the conjugate variable to $\rho$.  I now absorb the dependence on $M_p$ by introducing a dimensionless time $t' = t / \tau_{Sec}$ with
\begin{equation} \label{tau}
\tau_{Sec} = \frac{M_\star}{M_p}P,
\end{equation}
where $M_\star$ is the mass of the central star, $1.92 \pm 0.02 M_\odot$ \citep{Mamajek12}, and P is the test-particle's orbital period (which is constant since it is given by the test particle's semimajor axis through Kepler's third law).  This yields,
\begin{equation}
\frac{1}{P} \frac{d\rho}{dt'} = \pm M_\star \frac{\partial \mathcal{H}'_p}{\partial \sigma},
\end{equation}
which is independent of $M_p$.  This means that if a given $M_p$ yields a given orbital history, one can immediately generate an orbital history for a different $M_p'$ by stretching the time axis by the ratio $M_p/M_p'$.  This is a familiar feature of a test particle undergoing Kozai oscillations, which is a specialized case of secular dynamics with a single perturber.  

Since the test particles have no mass with which to affect Fomalhaut b, the latter's orbit remains fixed throughout the simulation.  I therefore construct my coordinate system around Fomalhaut b's fixed path.  I first choose the planet's orbital plane as the reference plane.  Thus, Fomalhaut b's orbital inclination $i_p = 0$.  Furthermore, I choose the direction toward the planetary orbit's pericenter as the $x$ axis, which sets the planetary orbit's longitude of pericenter to 0.  Thus, the only remaining (constant) parameters, are Fomalhaut b's orbital semimajor axis $a_p$, and eccentricity $e_p$.  I sampled $e_p$ over the range $[0.05,0.95]$ in increments of $0.1$, and $a_p$ from 60 to 340 AU in steps of 20 AU, the latter range covering approximately two standard deviations around the best-fit value of $a_p$ found by K13.  I thus performed 150 separate simulations for each possible combination of $a_p$ and $e_p$.  

Finally, for each simulation, I must choose the initial orbital elements for the test particles.  I assign particles orbital eccentricities of 0.1, the approximate value observed for the debris disk \citep{Kalas05}, and the semimajor axis corresponding to the belt's brightness peak of 142 AU (K13).   I take only a single test-particle orbital semimajor axis and vary the planet's $a_p$ since, as mentioned above, the dynamics only depend on the ratio of these two values.  One can therefore take my final results and translate them to other values of $a$ by adequately scaling $a_p$.  

I sampled the test-particle orbital angles as follows.  K13 find that, to $95\%$ confidence, the disk's inclination ($i$) relative to Fomalhaut b's orbital plane is under $45^\circ$ (see Fig.\ 22 in their work).  K13 also find that $90\%$ of the orbits that can fit the data are apsidally aligned with the disk to within $36^\circ$ (see Fig.\ 23 in their manuscript).  However, partly in order to make my results applicable to the general problem of radially crossing orbits, I chose to independently sample both the longitude of the ascending node ($\Omega$) and the argument of pericenter ($\omega$).  Of course, the apsidally aligned case still is a subset of my initial conditions.  In each simulation, I generated one thousand initial conditions for all combinations of ten equally spaced values of $i$ in the range $[0,45^\circ]$, and ten equally-spaced values for both $\Omega$ and $\omega$ in the range $[0,360^\circ]$. 

I performed all the numerical integrations with the secular integrator smpGauss \citep{Touma09}.  A numerical difficulty for integrating radially overlapping orbits is that as the orbits precess, two orbits can intersect at a point.  In such an event, the mutual forces suffer a discontinuity that is difficult for integrators to handle.  The integrator smpGauss circumvents this difficulty by using softened gravity.  This approximation reflects the reality that collisions are unlikely even in cases where two orbits intersect at a point, since the probability that both objects will cross the intersection point at the same time is generally small.  One would therefore expect that the softened secular solution should match a full integration until the test-particle suffers a close encounter with the planet.  I used a softening length of 0.01 AU, and verified that the secular integration matched a full N-body integration to within $\sim 1\%$, though only for about one secular timescale $\tau_{Sec}$ (cf. Eq.\ \ref{tau}).  I found that this is due to the secular dynamics being chaotic for overlapping orbits, with neighboring trajectories diverging after $t \sim \tau_{Sec}$.  This limitation is acceptable, since below I only consider a single secular period to constrain Fomalhaut b's properties.

I then performed the numerical integrations for the described suite of initial conditions, searching for regions of $(a_p, e_p)$ parameter space in which the particles' orbital eccentricities remained low on secular timescales.  The observed belt has a width $\sim 20 AU$, or $\sim 15\%$ of its semimajor axis.  In principle, this could be explained through parent bodies with a single semimajor axis and an eccentricity dispersion $\sim 0.15$.  To be conservative, I take twice this value ($e = 0.3$) as a threshold beyond which eccentricities are incompatible with observations.  Figure \ref{percents} shows the percentage of the thousand initial conditions that stayed below $e = 0.3$ over a single secular cycle in each simulation and are therefore nominally consistent with observations.  The solid white lines are the boundaries of the region of parameter space where Fomalhaut b's orbit would radially overlap the test particle's path if one approximated the latter orbit as circular.  Above these lines, Fomalhaut b's orbit is eccentric enough to cross the test particle's path.  

To within the resolution of the grid, all initial conditions are driven to high eccentricities in configurations where Fomalhaut b overlaps the test-particle's orbit (above the two curved solid white lines).  This result is general---a massive body on an eccentric orbit that crosses a belt of test particles on near-circular orbits will globally drive the latter onto similarly elliptical paths on secular timescales (not just particles that suffer close encounters).  This agrees with the intuition from low eccentricity and inclination Laplace-Lagrange secular theory; however, the result was not clear {\it a priori} since a) the high eccentricities and inclinations mean that additional terms in the disturbing function are important \citep[e.g.,][]{Murray99}, and b) the disturbing function's expansion is itself predicated on the assumption that the orbits do not cross.

The dashed white lines in Fig. \ref{percents} bound the region of phase space that is consistent with the observed deprojected distance of Fomalhaut b from the central star of 119 AU \citep{Kalas08}; below these lines, Fomalhaut b orbits too close (far) to reach its observed position at apocenter (pericenter).  The cross represents the best-fit value of $(a_p, e_p)$ from K13, with their marginalized 1-$\sigma$ error bars.  The figure shows that the region of phase space where Fomalhaut b's orbit is consistent with its observed distance from the star (above the dashed white lines) and does not drive parent bodies onto high-eccentricity orbits (lighter shades in grayscale) is narrow, restricted to small eccentricities, and inconsistent with the orbit determined by K13 (solid white cross)---see also Fig. \ref{times}.

I note that one cannot argue that we are observing the belt at a low eccentricity as part of a long-lived, larger-amplitude cycle.  As mentioned previously, the secular evolution is chaotic on a timescale $\sim \tau_{Sec}$; we would therefore see different belt particles at widely varying phases along the cycle, forming a much more radially extended disk than is observed.  

The evolution is also rapid compared to the age of the system (440 Myr).  Fig.\ \ref {times} shows the same grid as Fig.\ \ref{percents}, but assumes that Fomalhaut b has a mass equal to Saturn's, and the color scale represents the median time required for parent bodies in the given grid point to reach $e=0.3$ (out of the initial conditions that did so).  Only grid points where more than $90\%$ of the initial conditions were driven to high eccentricities (see Fig.\ \ref{percents}) were coded in this way---other grid points were colored white to indicate they may be dynamically stable.  The times in the figure can be immediately translated to a different-mass Fomalhaut b by multiplying the values by the ratio of Saturn's mass to the mass of interest (see Eq.\ \ref{tau}).  One can also interpret Fig.\ \ref{times} as the rough times Fomalhaut b would require to raise an initially circular debris disk to its present eccentricity of $e \approx 0.1$.  

\begin{figure}
\includegraphics[width=9cm]{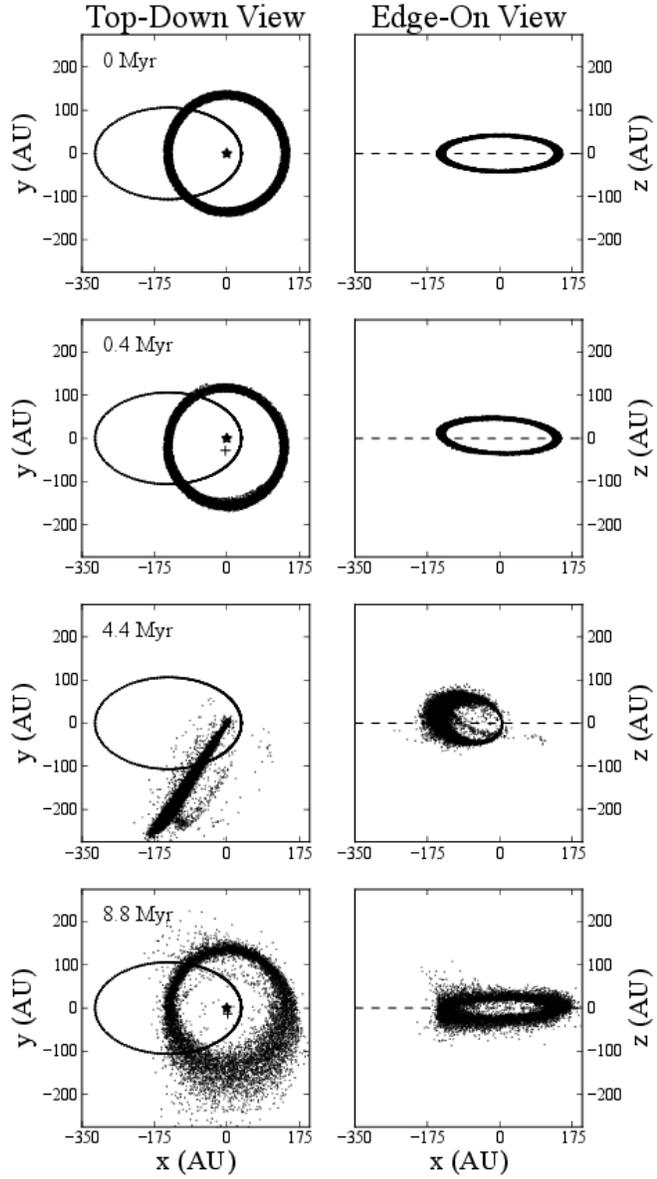}
\caption{\label{cycle}  Effect of Fomalhaut b (elliptical orbit plotted on left) on an initially near-circular, confined disk of massless parent bodies (with no additional perturbers).  Left panels show a top-down view, while the right panels show an edge-on view to Fomalhaut b's orbit---the dashed line represents Fomalhaut b's orbital plane.  The stars represent the central star, while the plus signs show the center of the debris disk, averaged over all test particles.  By t = 0.4 Myr (second row), the disk's geometry is consistent with that observed.}
\end{figure}

An important additional constraint is the observed alignment between the pericenters of the debris disk and Fomalhaut b.  It is also unclear whether the disk would retain a coherent shape as the eccentricities rise to the observed value of $e \approx 0.11$ \citep{Kalas05}.  To test this, I created a massless debris disk of $10^4$ parent bodies that is initially centered at 142 AU, 20 AU.  I assumed the initial random velocities in the debris were low and isotropic, drawing initial eccentricities and inclinations from uniform distributions over $[0,0.01]$ (with inclinations in radians).  I then gave Fomalhaut b the mass of Saturn, and took K13's marginalized best-fit elements of an orbital semimajor axis of 177 AU, eccentricity 0.8 and inclination to the disk of $17^{\circ}$.  I took the mutual line of nodes to coincide with Fomalhaut b's major axis---this orientation is difficult to determine observationally since the mutual inclination is low.  Snapshots of the evolution are shown in Fig.\ \ref{cycle}.  The left panels show a top-down view (looking down Fomalhaut b's orbit normal), while the right panels show an edge-on view to Fomalhaut b's orbit---the dashed line represents Fomalhaut b's orbital plane.  The stars represent the central star, while the plus signs show the instantaneous center of the debris disk, averaged over all test particles.  

By t = 0.4 Myr (second row), the parent-body orbits have evolved in near-unison to eccentricities of 0.14, comparable to the mean belt eccentricity measured by \cite{Kalas05} of 0.11.  This can be seen in the offset (plus sign) in the center of the debris disk from the central star (a low-eccentricity orbit is an offset circle to lowest order).  The longitudes of pericenter, measured relative to Fomalhaut b's pericenter are narrowly distributed around $84.7^{\circ}$ (standard deviation is $2.5^{\circ}$).  The eccentricities then continue to rise, accompanied by a dramatic secular inclination evolution.  By t = 4.4 Myr (third row), the orbits have rotated about their respective major axes into nearly polar orbits, and the eccentricities have reached their maximum value of $\approx 0.95$.  The orbits then continue rotating about their major axes, rendering them retrograde.  By the time the eccentricities decrease once again to values $\sim 0.11$ (t = 8.8 Myr, bottom row), the mean inclination of the disk-particle orbits to Fomalhaut b's orbital plane is similar to the corresponding epoch at 0.4 Myr (second row), except in a {\it retrograde} sense.  However, because the orbital evolution is chaotic on these secular timescales, the particle orbits have by this time diverged from one another, rendering the disk both radially and vertically extended, inconsistent with observations.  Over longer timescales than those plotted, particle orbits diverge further, and continue flipping between prograde and retrograde orbits.  Such flips are qualitatively very similar to those analyzed by \cite{Li13} while studying the effect of an outer eccentric perturber on an inner test particle that is also on an eccentric orbit; however, their analytical work cannot be directly applied to the present case where the orbits cross.  

I tested the robustness of the above results by trying seven other equally spaced orientations for the line of nodes---in all cases, similarly eccentric disks were formed at 0.4 Myr, the mean longitude of pericenter was within ten degrees of $90^{\circ}$, and orbits flipped back and forth between prograde and retrograde senses.  This behavior is consistent with that reported first by \cite{Beust13} (see note following the acknowledgements).

This predicted near-right angle between the disk's and Fomalhaut b's pericenters is largely ruled out by the data; however, since fits to the data with such large misalignments do exist (see the green points at near-right angles to the disk's major axis in Fig. 24 of K13), I briefly consider the consequences of such a scenario. 

\begin{figure}
\includegraphics[width=9cm]{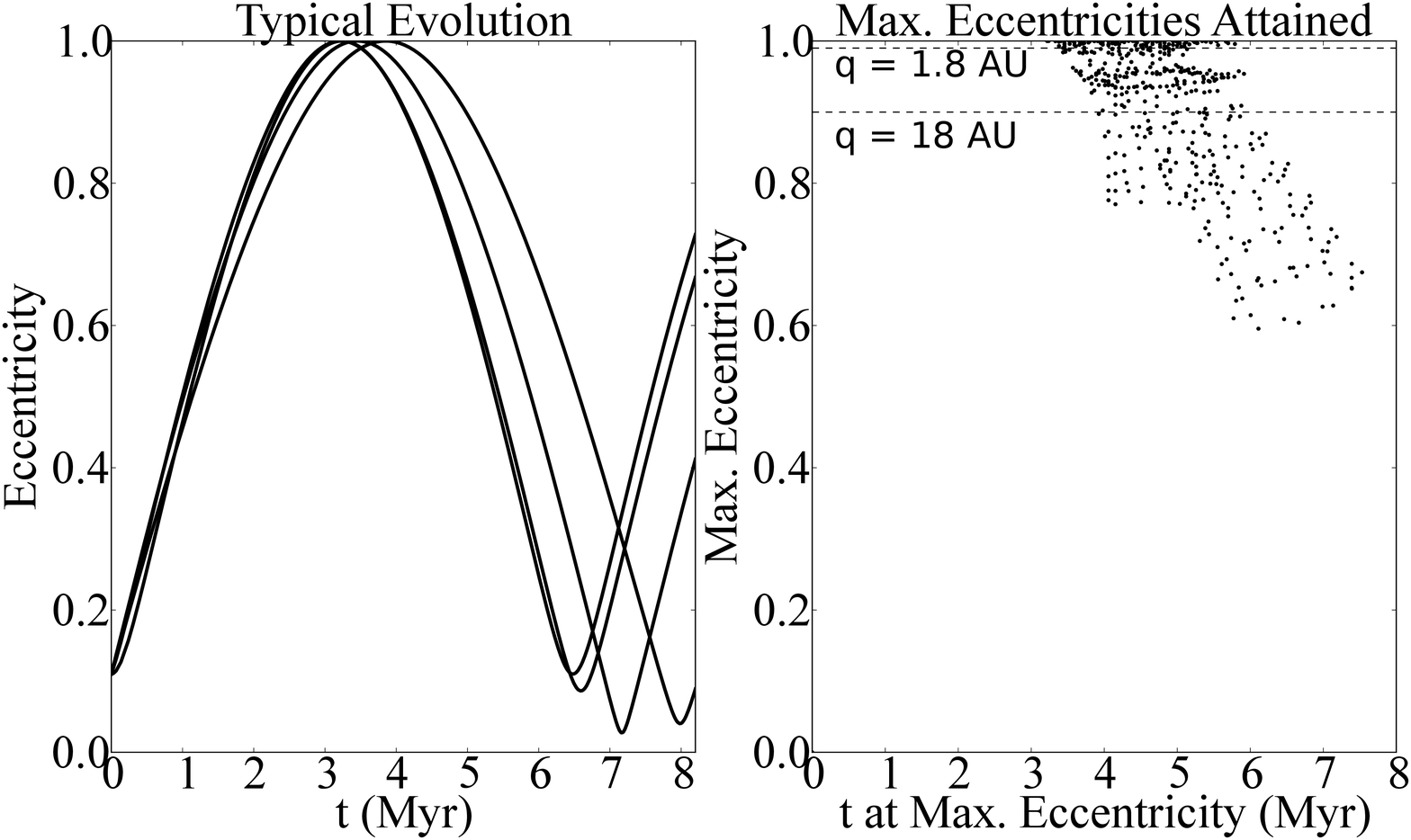}
\caption{\label{figLHB}  Left panel:  Secular evolution of four debris disk particles with similar initial conditions, drawn from the grid point closest to Fomalhaut b's best-fit orbit from K13 in Fig.\ \ref{percents} (a=180 AU, e=0.85).  All four particles reach eccentricities near unity, and begin to diverge after $\sim 1$ secular oscillation as mentioned in Sec.\ \ref{notOnlyPlanet}.  The simulation assumes Fomalhaut b has the mass of Saturn.  Right panel:  Maximum eccentricity attained by each of the thousand initial conditions for the same grid point mentioned above, plotted vs. the time at which the maximum value was achieved.  The dashed lines mark pericenter distances of $q = 1.8$ and $q = 18$ AU, showing that a large fraction of objects will reach the innermost Fomalhaut system.  }
\end{figure}

Figure \ref{figLHB} shows the secular evolution of four debris disk particles with similar initial conditions, drawn from the grid point in Fig.\ \ref{percents} closest to Fomalhaut b's best-fit orbit from K13 (a=180 AU, e=0.85).  It assumes Fomalhaut b has the mass of Saturn, but one can simply scale the time for a different mass $M_p$ by multiplying all the times by the factor $M_{Sat}/M_p$ (see Eq.\ \ref{tau}).  

The right panel of Fig.\ \ref{figLHB} shows the maximum eccentricities attained in the first secular cycle for the thousand initial conditions in the same grid point of Fig.\ \ref{percents} identified above.  The dashed lines represent pericenter distances of 1.8 and 18 AU.  I find that 66$\%$ of initial conditions drop below 18 AU, and $30\%$ fall below 1.8 AU.  These values vary by about $15\%$ between adjacent grid points.  Thus, within the next few Myr, the inner Fomalhaut system would be injected with a large number of high-eccentricity impactors.  If the Fomalhaut system harbors interior planets, they would then undergo a violent period perhaps analogous to the Solar System's hypothesized Late Heavy Bombardment 4.1-3.8 Gyr ago, when the impact rate on the Moon and other terrestrial planets may have spiked \citep{Hartmann00}.

The main conclusion of this section is that it is unlikely that Fomalhaut b is a Neptune-mass or larger planet if it is the only body interacting with the debris disk.  Not only would Fomalhaut b need to have scattered into its present orbit in the last $\sim 1-10$ Myr to not have disrupted the disk, current data largely rules out the predicted $\sim 90^{\circ}$ misalignment between the disk's and planet's pericenters.  On the other hand, the deleterious gravitational effects from a Fomalhaut b roughly less massive than Neptune would be attenuated by the disk's self-gravity.  In this case, however, one must invoke an alternate mechanism for producing the observed eccentricity of the debris disk.  I thus turn in the next two sections to the alternate possibility that an unseen planet is dominantly responsible for forcing the debris disk's observed eccentricity and shaping its sharp inner edge.  
 
\section{If an unseen planet dominantly forces the eccentricity in the debris disk, what is Fomalhaut b's maximum lifetime on its present orbit?} \label{fomc}

I argued in the introduction that if Fomalhaut b cannot force the debris disk's observed eccentricity (because it is either too light or scattered into its present orbit too recently), then the best explanation for the belt's elliptical geometry is an unseen planet (Fomalhaut c) orbiting interior to the disk.  Importantly, the addition of a second planet relaxes the constraints from the previous section.  If Fomalhaut c's mass ($M_c$) is larger than that of Fomalhaut b ($M_b$), the former can dominate the secular dynamics and test particles can be partially shielded from the latter's extreme effects.  Nevertheless, the fact that we observe the belt as dynamically cold sets important limits on Fomalhaut b's mass and on its lifetime in its current orbit.  

My numerical experiments, detailed below, are similar to ones performed by K13 (see their Sec.\ 9.3.3).  The key difference is one of timescale.  K13 were interested in showing that physical crossings of Fomalhaut b through the disk would not destroy the belt structure.  They therefore chose the most damaging configuration---namely, a coplanar system where Fomalhaut b plows through the disk every orbit.  They found that, indeed, Fomalhaut b does not interact with enough of the disk's circumference to erase its structure for $M_b \lesssim 1 M_J$.  However, the coplanar arrangement limited their integrations to less than 0.5 Myr, since eventually Fomalhaut b and Fomalhaut c would scatter and the system would no longer be coplanar.  But these timescales are too short to probe the dramatic secular effects we found in Sec.\ \ref{notOnlyPlanet}.  I therefore simulate longer times in a non-coplanar configuration, consistent with the observation that Fomalhaut b's orbit is likely mutually inclined with the disk (K13).  

Perhaps the most constraining feature of the debris disk is its narrow vertical extent.  The radial extent of the belt could support a comparatively wide range of semimajor axes and eccentricities, but the relative inclinations between particles must unambiguously remain small.  Photometric models by \cite{Kalas05} give an opening angle of $\approx 1.5^{\circ}$.  For $M_b \ll M_c$, one expects test particles to remain coplanar with Fomalhaut c, but as $M_b$ approaches $M_c$, the inclined Fomalhaut b (marginalized best-fit value $i \approx 17^\circ \pm 12^\circ$, see Fig.\ 22 in K13) should pull particles out of the plane on secular timescales.  The conclusions from this section should hold as long as Fomalhaut b's inclination to the debris disk is $\gtrsim$ the disk's opening angle.

Since these secular timescales are shorter than the age of the system for $M_b \gtrsim 1 M\oplus$, we must once again conclude that Fomalhaut b scattered onto its present path relatively recently.  I artificially simulate this event by injecting Fomalhaut b into its current orbit only after letting Fomalhaut c dynamically shape the disk's inner edge and allowing the belt to reach a quasi-steady state.  I then recorded the time required for Fomalhaut b to disrupt the disk, as a function of planet mass.  Infrared surveys to date that have failed to detect planets in the Fomalhaut system have been sensitive to planet masses $> 1 M_J$ \citep{Kalas08, Marengo09, Janson12}.  I therefore assign Fomalhaut c a mass of $1 M_J$---the best-case scenario for shielding the belt from Fomalhaut b and thus for system stability.  I will argue that Fomalhaut b's lifetime on its present orbit must be much shorter than the system's age (making it unlikely to observe the planet at this special time), so lower Fomalhaut c masses would only exacerbate the problem.  The model is purposefully simple since there are currently too many unknowns to perform an in-depth dynamical study spanning the space of possible initial conditions.

\begin{figure}
\includegraphics[width=9cm]{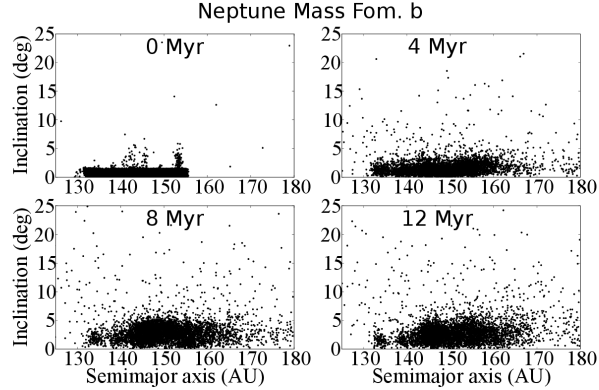}
\caption{\label{snaps} Snapshots in time of the inclination distribution of parent bodies vs. semimajor axis.  The hypothetical unseen planet Fomalhaut c has a mass of 1 Jupiter-mass, and Fomalhaut b has a mass of 1 Neptune-mass.  By $\approx 10$ Myr, the mean inclination has changed, and the inclination dispersion has doubled from its initial value, inconsistent with observations.}
\end{figure}

\begin{figure}
\includegraphics[width=9cm]{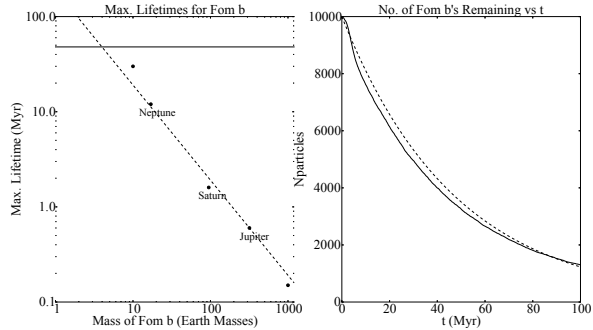}
\caption{\label{lifetimes}  Left panel:  Summary of simulations with a Jupiter-mass Fomalhaut c.  The plot shows Fomalhaut b's maximum lifetime on its present orbit (defined as the time required for the planet to cause the debris disk's inclination dispersion to double), as a function of Fomalhaut b's mass.  The dashed line shows the expected scaling $\propto M_p^{-1}$ that one expects from secular behavior, normalized to the maximum lifetime of Jupiter.  Planets in the limit $M_p \ll M_c$ are ejected on a timescale $\sim \tau_{Scat} =$ 48 Myr, shown by the solid horizontal line.  Right panel:  Number of initial Fomalhaut b orbits surviving vs. time (solid line).  The trend is well fit by an exponential decay with $\tau_{Scat} = 48$ Myr (dashed line).  On a timescale $\tau_{Scat}$, objects suffer a close encounter with Fomalhaut c and are ejected from the system.  See text for initial conditions.}
\end{figure}

My detailed initial conditions were as follows (results are in Figs. \ref{snaps} and \ref{lifetimes}).  I take Fomalhaut c to be on the orbit that \cite{Chiang09} found to best fit the disk's morphology for a $1 M_J$ planet, and Fomalhaut b to be approximately on the best-fit orbit found by K13.  Specifically, for Fomalhaut c, I adopted $a_c = 109 AU$ and $e_c = 0.123$, where the c subscript corresponds to Fomalhaut c \citep{Chiang09}\footnote{Subsequent to the work of \cite{Chiang09}, Fomalhaut A's mass has been revised downward by $\approx 20 \%$ by \cite{Mamajek12}.  Thus, I am using elements that would correspond to a $1.2 M_J$ planet.  This is a minor effect, especially given my other uncertainties, since the planet mass' main effect is in setting the location of the disk's inner edge, and I determine this numerically by integrating test particles for $100$ Myr prior to introducing Fomalhaut b.}.  I reference the coordinate system to Fomalhaut c's {\it initial} orbit (which will change once Fomalhaut b is introduced), so at first $i_c = 0$ and the longitude of pericenter is 0.   I then introduce $10^4$ test particles representing the parent bodies in the debris disk.  For each orbit, I randomly assign $125 < a < 155$ AU, $0^{\circ} < i < 1.5^{\circ}$, $0^{\circ} < \Omega < 360^{\circ}$, and $\omega = -\Omega$ so that the orbits are apsidally aligned with Fomalhaut c (the longitude of pericenter $\varpi = \Omega + \omega = 0$).  The orbital eccentricities were set to the forced values calculated from Laplace-Lagrange secular theory \citep[see Eq.\ 13 in][]{Chiang09}.  I chose the semimajor axis range to roughly match the observed disk's radial extent, though it extends further inward so that I can numerically determine the boundary of the planet's chaotic zone where particles should be depleted.  The inclination range was chosen to match the photometric models of \cite{Kalas05}. 

I then integrated Fomalhaut c and the $10^4$ test particles for 100 Myr, in order to numerically establish the inner boundary of the planet's chaotic zone and establish realistic initial conditions.  This follows the approach taken by \cite{Chiang09}, and I find that the 100-Myr timescale is sufficient for the system to reach a quasi-steady state where the particle population is approximately constant.  For the integrations I used the package SWIFTER, written by D.E. Kaufmann, which is based on SWIFT \citep{Levison94}.  I chose to use the integration scheme SyMBA within SWIFTER \citep{Duncan98, Levison00}, as it allowed me to accurately integrate close encounters with Fomalhaut c while ignoring interactions among test particles.  Particles were removed from the simulation if they physically collided with Fomalhaut c, or if they moved inside 0.008 AU or outside 1000 AU.  After 100 Myr, 7433 particles remained, and, like \cite{Chiang09}, I find an inner edge at approximately 131 AU, with inclinations largely unchanged, except at locations corresponding to mean-motion resonances with Fomalhaut c. 

With this quasi-steady state of initial conditions, I then introduced Fomalhaut b on its extreme orbit.  To be consistent with K13's best-fit orbit, I chose an apsidally aligned configuration ($\Omega = 0$, $\omega = 0$) with $a = 177$ AU, $e = 0.8$, $i=15^{\circ}$.  The $a$ and $e$ values are the marginalized values found in the same study, and $i$ is slightly smaller than the best-fit value of the mutual inclination ($17^{\circ}$).  

Figure \ref{snaps} shows snapshots in time of the inclination distribution for a Neptune-mass Fomalhaut b, which confirm the expectations presented at the beginning of this section.  Initially (following the 100 Myr integration without Fomalhaut b), the standard deviation of the parent-bodies' orbital inclinations was $3.4^{\circ}$ (biased beyond $1.5^{\circ}$ by inclinations excited at mean-motion resonances as mentioned above---many points lie off-screen at higher inclinations).  By 10 Myr, the mean inclination has risen, the standard deviation has doubled, and the distribution is clearly inconsistent with observations.

The main result of this section is the left panel in Fig.\ \ref{lifetimes}, which shows a summary of the results of our simulations where Fomalhaut b has masses of either 1 $M_\oplus$, 3 $M_\oplus$, 10 $M_\oplus$, Neptune (17$M_\oplus$), Saturn (95 $M_\oplus$), Jupiter (318 $M_\oplus$), or 1000 $M_\oplus$.  I took the maximum lifetime of Fomalhaut b on its present orbit to be the time required for the standard deviation of the inclinations to double (cf. Fig.\ \ref{snaps}).  The dashed line (normalized to the value for Jupiter) shows the expected scaling $\propto M_p^{-1}$, where $M_p$ is the planet's mass, that one expects from secular behavior (see Eq.\\ \ref{tau}).  The trend may have continued for slightly smaller values than $M_p = 10 M_\oplus$, but in my runs with $1 M_\oplus$ and $3 M_\oplus$, the integration was long enough that Fomalhaut b suffered a close encounter with Fomalhaut c and was ejected from the system.  Thus, for objects with $M_p \ll M_c$, the maximum lifetime is set by the close-encounter timescale with Fomalhaut c.  

To quantify this ejection time in the limit $M_p \ll M_c$, I performed an integration using SyMBA \citep{Duncan98} with the same Fomalhaut c, and $10^4$ hypothetical Fomalhaut b test particles that do not interact with each other.  These candidate planets were given the same $a = 177$ AU, $e = 0.8$ and $i=15^{\circ}$, but their longitudes of ascending node, arguments of pericenter and true anomalies were randomly drawn from a uniform distribution in the range $[0,360]$.  The solid line in the right panel of Fig.\ \ref{lifetimes} shows the number of surviving Fomalhaut b's vs. time.  It is well-fit by an exponential decay (dashed line) with an e-folding timescale of $\approx 48$ Myr.  This value is plotted as the solid horizontal line in the left panel.  

So while K13 found that a Neptune-to-Saturn-mass Fomalhaut b would not disrupt the disk on orbital timescales (see their Fig.\ 29), I conclude that such a planet must have scattered onto its present orbit more recently than the secular timescale of $\sim 10$ Myr ago.  This is essentially the same result I obtained in the previous section's scenario where Fomalhaut b was the only object interacting with the disk.  It should not be surprising as they are both measuring the secular timescale of perturbations from Fomalhaut b.  Sub-Neptune Fomalhaut b masses would also be consistent, since then the disk's self-gravity (which I have ignored) would become important, and Fomalhaut b's secular effects on the belt are mitigated.  However, on a timescale $\tau_{Scat} \approx 48$ Myr, Fomalhaut b would undergo a close encounter with Fomalhaut c and be ejected.  This constraint is relaxed with lower masses for Fomalhaut c, since closer encounters become necessary for ejection, so $\tau_{Scat}$ rises.  For example, in a parallel simulation to that displayed in the right panel of Fig.\ \ref{lifetimes} using a Saturn-mass Fomalhaut c, I find $\tau_{Scat}$ = 194 Myr.  

I conclude that if there exists an unseen Fomalhaut c orbiting interior to the debris disk and forcing its observed eccentricity as described by \cite{Quillen06} and \cite{Chiang09}, then either (a) Fomalhaut b is massive enough ($\gtrsim$ Neptune-mass) to dominate the disk's self-gravity and must have scattered into its present orbit within the last $\sim 10$ Myr, or (b) Fomalhaut b's mass is small relative to the debris disks's and can thus be long-lived without destroying the disk's structure.  The latter case is more likely for Fomalhaut c masses $\sim$ 1 Saturn-mass, as then the timescale for close encounters between Fomalhaut b and c leading to ejection of Fomalhaut b is comparable to the age of the system.  But if one considers the possibility that Fomalhaut b's mass is low enough that its secular effect on the debris disk is negligible, one should also investigate other low-mass scenarios---perhaps Fomalhaut b is nothing more than an isolated, transient dust cloud.  I consider this general class of low-mass scenarios for Fomalhaut b in the next section.

\section{A Low-Mass Fomalhaut b?} \label{dust}

Before considering specific possibilities, I begin by making two general points about any models that claim that Fomalhaut b's mass is low enough ($\lesssim$ Neptune-mass) that it does not influence the secular dynamics in the debris disk.  First, by emasculating Fomalhaut b, one pushes the burden of explanation for the circumstellar disk's measured eccentricity onto an unidentified perturbation, such as an unseen planet.  Second, it is difficult to explain why Fomalhaut b and the disk seem to be roughly apsidally aligned (to within at least $\sim 25^{\circ}$, K13).  By the above definition, a low-mass Fomalhaut  b cannot be forcing the alignment; it is perhaps possible that instead the disk's gravity forces Fomalhaut b's pericenter, though one would also have to consider the secular effects from Fomalhaut c (or whatever other perturbation is responsible for the disk's forced eccentricity).  I leave such an analysis for future work when more data is available and the problem is better constrained.  Of course, the alignment could also be purely fortuitous (probability $\lesssim 10\%$).

The next important constraint for a low-mass Fomalhaut b is the object's large observed flux at optical wavelengths.  Because Fomalhaut b has not been detected in the infrared, optical images must be detecting starlight scattered from a vast cloud of dust.  One can envision a variety of explanations for such dust, which I deal with in turn.  In all cases, the main question is whether such a dusty structure would be long-lived enough, or such dust-producing events frequent enough, to render it plausible that we would observe Fomalhaut b today.

By assuming the dust is optically thick, \cite{Kalas08} estimated the minimum-size disk of dust that could explain the observed flux as $\sim 300$ Earth-radii.  While a primordial protosatellite disk of this size would be plausible around a gas giant during planet formation, it is unlikely that a super-Earth would host such a disk 440 Myr after formation.  Circumplanetary disks from giant impacts like the ones thought to have formed the Moon \citep[e.g.,][]{Salmon12} and Charon \citep[e.g.][]{Canup11} are also highly implausible, since such disks are much smaller than 300 Earth-radii and should only last $\sim 100$ yrs \citep{Salmon12}.  

A more promising idea proposed by \cite{Kennedy11} is that we are observing dust produced in a collisional cascade among a swarm of irregular satellites around Fomalhaut b.  Such populations are observed around each of the Solar System's giant planets, and fill a large fraction of their parent planets' Hill spheres (inside which the planet can keep satellites in orbit in the face of the Sun's tidal gravity).  Such a dust distribution would be more extended and optically thin, though still unresolvable by HST, since the Hill radius of a 2 $M_\oplus$ Fomalhaut b on the nominal orbit from K13 would be $\approx 0.4$ AU and the Hubble Space Telescope's ACS/HRC has a point spread function with a full-width half-max corresponding to $\approx 0.5$AU \citep{Kalas08}.  Note that to estimate Fomalhaut b's Hill sphere on its eccentric orbit, it is important to evaluate the Hill radius at pericenter \citep[e.g.,][]{Hamilton92}. Using a simple model, \cite{Kennedy11} argue that the minimum planet size that can host such a collisional cascade of irregular satellites is set by the requirement that the orbital speeds be fast enough for collisions between irregular satellites to be destructive.  They find that for the collisional cascade to generate the observed flux for the age of the Fomalhaut system, Fomalhaut b must have a mass $\gtrsim$ a few $M_\oplus$.  Thus, subject to the caveats mentioned at the beginning of the section for all low-mass Fomalhaut b scenarios, this provides a possible, though finely tuned explanation:  Fomalhaut b is a super-Earth with enough mass to drive a collisional cascade of irregular satellites (M $\gtrsim$ a few $M_\oplus$) but too little to drive the secular dynamics of the debris disk (M $\lesssim$ Neptune-mass).  

However, once one posits that Fomalhaut b is of such low mass that its dynamical influence on the system is negligible, it is tempting to throw out the planetary interpretation completely and argue that Fomalhaut b is simply a transient dust cloud.  Such a scenario was considered by \cite{Galicher13}, but the revised eccentric orbit of Fomalhaut b provides new constraints.

One might ask whether a cloud of dust could naturally be driven to high eccentricity through radiation pressure.  This is appealing as it could in principle account for Fomalhaut b's anomalously high eccentricity through a collision between objects on near-circular orbits.  Objects moving on such paths should be much more numerous (particularly in the circumstellar debris disk), so one would expect many more such collisions compared to ones involving eccentric orbits.  However, on short sub-orbital and size-variable timescales, radiation pressure would disperse such a dust cloud to several AU in extent \citep{Burns79}.  Fomalhaut b would have thus been easily resolved in the original HST observations, which it was not \citep{Kalas08}.  \cite{Galicher13} argue that Fomalhaut b may be marginally resolved in the same HST data, though the inferred size of the dust cloud $\approx 0.5$ AU is still far too small.  One might argue that the collision could have taken place recently (close to Fomalhaut b's current position) so that the dust would not have had time to disperse beyond $\sim 0.5$ AU.  But such a scenario also fails, as dust released from nearly circular orbits would move on orbits whose pericenters are close to their location of release \citep{Burns79}.  However, Fomalhaut b's pericenter is more than $90^{\circ}$ away from its present position (see Fig.\ 20 in K13).  Fomalhaut b's eccentricity is therefore not caused by radiation pressure; rather, it must be inherited from one (or both) of its parent bodies.  

The requirement that the putative Fomalhaut b dust cloud be smaller than 0.5 AU in radius (to be unresolved or marginally resolved) further constrains any impact event to have occurred recently.  If the resulting dust cloud is unbound to a massive central object, K13 find that Keplerian shear smears the cloud along the orbit into a structure that is resolvable by HST within $\approx 1000$ years.  K13 further determine that one can prevent such shearing by placing the cloud in orbit around a parent central body with mass $\gtrsim 5\times10^{21}$ kg (roughly 5 times Ceres' mass).  Perhaps, therefore, dust around a dwarf planet could render the configuration long-lived (without disturbing the disk, see Fig.\ \ref{lifetimes}).  However, as I argued above, such an impact-generated circumplanetary cloud like the one thought to have produced Pluto's moon Charon \cite{Canup11} would be far too compact and short-lived to explain Fomalhaut b.  Finally, unbound dust from such an impact would disperse even faster than in the isolated dust-cloud scenario.  The escape speed from objects $\gtrsim 5\times10^{21}$ kg is $gtrsim 0.5$ km/s.  Taking the unbound debris to collectively be moving at this minimum escape speed (this is really a three-body problem, but the two-body escape speed sets the right scale), or at $\gtrsim 0.1$ AU/yr, the cloud would be resolved by HST within $\lesssim 20$ years.

I conclude that since the lifetimes of clouds of debris that are either isolated or bound to low-mass objects are so short, the probability of observing a specific such dust cloud is exceedingly low.  However, one must balance this against the rates of collisions that produce such objects.  Previous estimates of these frequencies yielded reasonable values of $\sim 1$ dust cloud per 200 yrs \citep{Galicher13}, but they assumed that the impactors originated in the densely populated debris disk.  This scenario must be re-evaluated in light of Fomalhaut b's revised eccentric orbit.  It is difficult to see from simple conservation laws how an impact between comparably sized bodies in the belt could simultaneously yield a low angular momentum (to provide a large eccentricity) and a high orbital energy (to raise $a$ from the disk's value of $\approx 140$ AU to Fomalhaut b's $a \sim 180$ AU).  For example, a prograde-retrograde collision between comparably sized objects could generate an appropriately low angular momentum, but would yield a low orbital energy and thus a {\it smaller} semimajor axis.  One seems to require the rarer event of an object first scattering onto Fomalhaut b's orbit, and then impacting a much smaller body in the debris disk so that the total energy and angular momentum is dominated by the former projectile; the resulting dust cloud would then follow the same eccentric orbit.  

I briefly note that if the potential scattered object does not cross the debris disk, the collision probability plummets.  The particle-in-a-box impact timescale between two isolated bodies is $\sim (a/R)^2$ orbital times, where $a$ is the semimajor axis and $R$ is the impactor radius \citep[e.g.,][]{Hamilton94}.  The observed flux from Fomalhaut b corresponds to a dust mass equivalent to that in a $\sim 10$ km body \citep{Kalas08}.  If one considers a collision between two such objects with semimajor axes $\sim 100$ AU, one obtains an impact timescale of $\sim 10^{21}$ yrs.  Even if there were many such candidate pairs of objects to collide, this is prohibitively long.

In conclusion, the only long-lived explanation for a low-mass Fomalhaut b considered above is a collisionally grinding swarm of irregular satellites around a super-Earth with $\sim$ a few $M_\oplus \lesssim M \lesssim$ Neptune-mass.  Other impact scenarios all have short dust lifetimes $\lesssim 1000$ yrs.  This can be counterbalanced if the frequency of such impacts is large.  I argued that plausible impacts require an original impactor on a similarly eccentric orbit that physically crossed through the debris disk (not just in projection).  Currently, only $\approx 12\%$ of the orbits that fit the data do so (K13), but this will be an important test as the orbit is further refined with additional observations.  This assumes that Fomalhaut b's orbit has not changed appreciably since the impact, which is reasonable since the dust-cloud lifetime of $\sim 1$ Fomalhaut-b orbit about the central star does not give the orbit time to change appreciably.  In both the long-lived irregular-satellite swarm and the frequent transient dust-cloud scenarios, we require the existence of an undetected giant planet Fomalhaut c orbiting interior to the observed debris disk (or a different undetermined mechanism) to force the belt's eccentricity.  An adequate explanation must also account for the seeming apsidal alignment between Fomalhaut b and the disk (K13).  

\section{Conclusion} \label{LHB}

In this article, I have considered three scenarios:  (1) Fomalhaut b is a giant planet and is the only object dynamically interacting with the debris disk, (2) Fomalhaut b has enough mass ($\gtrsim 1$ Neptune mass) to secularly affect the disk, but the belt's eccentricity is instead dominantly forced by an additional unseen planet Fomalhaut c, and (3) Fomalhaut b is a low-mass object---either a super-Earth generating dust through collisions between its irregular satellites, or perhaps simply a transient dust cloud---and again the belt's eccentricity is set by an unseen Fomalhaut c (or a different unidentified perturbation).  

Case (1) makes the important diagnostic prediction that the pericenters of Fomalhaut b and the debris disk should be at roughly right angles (Sec. \ref{notOnlyPlanet}, Fig.\ \ref{cycle}).  This was found recently by \cite{Beust13} (see the note following the acknowledgements).  Currently, such a scenario is largely inconsistent with observations.  Additional epochs of observation should be able to rule out the few orbital fits that are consistent with this scenario; however, in the event that such fits are instead supported by additional data, one expects disk particles to reach eccentricities approaching unity within $\sim 10$ Myr (Fig.\ \ref{figLHB}).  If the Fomalhaut system harbored planets close to the central star, such planets would undergo a period analogous to the hypothesized Late Heavy Bombardment in our own Solar System (Sec. \ref{notOnlyPlanet}).

In case (2), where the debris disk's observed eccentricity is instead dominantly forced by an unseen planet Fomalhaut c, a $\sim$ Neptune-mass or larger Fomalhaut b would have to have scattered into its present orbit within the last $\sim 10$ Myr (see Fig.\ \ref{lifetimes} in Sec. \ref{fomc}).  This timescale is short compared to the system's age of $440 $Myr, rendering it unlikely that we would witness the event in process.

Lower masses than $\sim 1$ Neptune-mass for Fomalhaut b (case 3) mean that Fomalhaut b's gravitational effect on the debris disk will be small compared to the disk's self-gravity, and the disk's structure can be maintained on long timescales.  However, such explanations require an undetected Fomalhaut c (or a different undetermined mechanism) to force the debris disk's observed eccentricity, and must simultaneously account for the large amount of dust implicated in Fomalhaut b's observed optical flux \citep{Kalas08}.

One plausible explanation is that the dust is maintained through a collisional cascade among a swarm of irregular satellites, which exist around each of the Solar System's giant planets.  However, \cite{Kennedy11} find that only planets $\gtrsim$ a few Earth masses have fast enough circumplanetary velocities for collisions between irregulars to be sufficiently destructive to maintain the dust over the Fomalhaut system's lifetime.  Thus, a planetary interpretation for Fomalhaut b only is plausible within a narrow range of masses between a few and $\sim 20$ Earth-masses.

Alternatively, Fomalhaut b could merely be a transient dust cloud, the result of one of many, frequent impacts.  I argued in Sec.\ \ref{dust} that Fomalhaut b's revised eccentric orbit requires an impact between a body previously scattered into Fomalhaut b's orbit that collides with a much smaller object.  It is otherwise difficult to explain Fomalhaut b's low angular momentum, high energy orbit.  I also argued that for such an impact to be plausible, it must have occurred recently, and Fomalhaut b's orbit must pass directly through the debris disk---currently only $\approx 12\%$ of the orbits that fit the data do so (K13).  It is also difficult under this scenario to understand the observed apsidal alignment between Fomalhaut b and the debris disk, since Fomalhaut b has no effect on the disk, and the dust cloud's lifetime ($\sim 1$ Fomalhaut b orbit) is not long enough for the disk or Fomalhaut c to secularly affect Fomalhaut b's orbit.  In this case, apsidal alignment would have to be the orientation under which Fomalhaut b's orbit intersects the debris disk, which the current data do not support. 

The Fomalhaut system thus provides a complex set of constraints on Fomalhaut b.  The best explanation found in this paper is that Fomalhaut b is a super-Earth hosting a large population of irregular satellites, and that there exists an undetected Fomalhaut c orbiting interior to the disk in the range $a \approx 100-120$ AU and $e \approx 0.11-0.14$---the orbit Fomalhaut b was previously thought to move on, which can naturally explain the debris disk's observed eccentricity \citep{Quillen06, Chiang09}.  A super-Earth mass is large enough to drive a collisional cascade among the irregular satellites to generate the observed dust, yet low enough so as to preserve the debris disk's structure on long timescales.   This narrow mass range is somewhat uncomfortable, but future observations should help elucidate Fomalhaut b's mysterious nature.  

One way to falsify this hypothesis would be to resolve Fomalhaut b's emission and obtain an estimate of the dust's extent.  The dynamically stable region in which irregular satellites can orbit over long timescales is inside roughly half of the planet's Hill sphere \citep[e.g.][]{Nesvorny03}.  This scenario then predicts that the dust emission should be confined to a cloud with radius $\lesssim 1\%$ of Fomalhaut b's pericenter distance $\sim 0.5 AU$.  This roughly corresponds to the resolution of the original images taken by \cite{Kalas08} with the Hubble Space Telescope.  Interestingly, \cite{Galicher13} find that Fomalhaut b may be resolved at 0.8 $\mu$m.  Further observations should help elucidate this picture.

In particular, JWST's NIRCAM should be in a powerful position to constrain the Fomalhaut system both by pushing to higher sensitivity and resolution in the infrared, and by extending the temporal baseline over which Fomalhaut b is tracked, thus refining its orbit.  The major questions are: (i) Can JWST detect Fomalhaut b in the near infrared, confirming it as a giant planet? (ii) Can it resolve Fomalhaut b, suggesting it is instead an expanding/shearing cloud of dust?  (iii) Can it find a Fomalhaut c orbiting interior to the disk (iv) Does Fomalhaut b's orbit physically cross through the debris disk, rendering it plausible for a collision between two low-mass objects to generate Fomalhaut b?

Of one thing we can be certain---as new observations come in, the Fomalhaut system will remain a rich and lively topic of debate.

\section{Acknowledgements}
I would like to thank Paul Kalas for alerting me of, and discussing with me, the interesting dynamical questions surrounding this system.  I am indebted to Joseph A. Burns for his careful reading of the manuscript, and would like to thank Philip D. Nicholson, Matthew H. Hedman, Matthew S. Tiscareno, Rebecca A. Harbison, and Steven H. Strogatz for insightful discussions.  I would also like to thank an anonymous reviewer who provided numerous insightful comments that greatly strengthened this manuscript.

{\it Note added}---As I was finalizing my response to the referee's comments on this manuscript, a similar, independent investigation of the system's dynamics by \cite{Beust13} was posted to the arXiv.  They undertook an independent reanalysis of the data, obtaining orbital fits consistent with those reported by K13, and performed a more detailed semi-analytic study of the scenario I investigated in Sec. \ref{notOnlyPlanet}.  This manuscript additionally performs numerical simulations involving an additional unseen planet Fomalhaut c (Sec.\ \ref{fomc}), and pursues implications for a low-mass Fomalhaut b (Sec.\ \ref{dust}).  


\end{document}